# Softening Theory of Matter: Tuning Atomic Border to Make Soft Materials

Sen Chen [1,2], Lei Wang [1], and Jing Liu [1,2,3*]

[1]Technical Institute of Physics and Chemistry, Chinese Academy of Sciences, Beijing 100190, China

[2]School of Future Technology, University of Chinese Academy of Sciences, Beijing 100049, China

[3]Department of Biomedical Engineering, School of Medicine, Tsinghua University, Beijing 100084, China

*Corresponding author. Email: jliu@mail.ipc.ac.cn

**Abstract:** Regulation of material's softness has both theoretical and practical significances due to irreplaceable applications of soft matter in rather diverse areas. This article is dedicated to draft a theoretical category on interpreting the mechanisms lying behind all soft matters, which can be termed as Softening Theory. Then a technical strategy with generalized purpose was proposed for softening desired matter, i.e. the melting point of matter can be significantly reduced through tuning its interior boundaries in atomic level. This theory accords well with the classical nuclear droplet model that treats the nucleus as a droplet which had in fact successfully explained many phenomena. It also explained the important experimental fact that, the material's melting point is often measured to be drastically reduced as the particles become smaller enough, in which situations effects of the atomic borders become much weaker. Along this direction, many phenomena existing in nature can be well understood. For example, if keeping in mind the fact that an atom consisting of nucleus and electronics can be regarded as fluid, all the matter consisted of atoms should maintain fluidic state in their macroscopic scale, according to the consistency between macro and micro worlds. However, many substances just cannot remain their original atomic fluidic behavior due to breaking of the consistency between macro and micro states. Based on the current softening theory, it is now easy to understand that the breaking of such consistency is just caused due to generated forces from the atomic interactions. To resolve such intrinsic confinement, a group of potential technical approaches can be developed such as via mechanical, electrical, thermal, optical energy including alloy processing etc. to tune the atomic borders of the matter and thus make desired soft materials. This work provides a theoretical foundation to partially address the nature of the material which will aid to make future soft matters in the coming time.

## 1. Introduction

Creatures that are composed of soft tissues often show excellent flexibility. In fact, ubiquitous softness plays an important role in life activities, as shown in Figure 1. In a large extent, softness can be regarded as another dimension outside the geometric or time scales to characterize matter. Clearly, it is of significance to regulate the softness of matter. Especially, recent interest in metallic materials with a low melting point near room temperature has increased due to their potential use in chip cooling, soft robot and biomedical practices, etc. Unfortunately, most classical metals possess a very high melting point due to presence of the strong metal bonds. Therefore, it is of physical and practical significance to reduce the melting point of the material. A softening theory for matter was thus introduced here which suggests a new strategy for tuning the melting point of the target matter.

One important fact on matter is that, as the metal particles become smaller, the metal's melting point is drastically reduced *(1-3)*. This phenomenon is well known in academics and many theories have been proposed to explain it *(4) (5)*. However, the available interpretations so far are still not ideal enough. From an alternative, here we would suggest another different perspective to describe such phenomenon. We point out that particle can become a droplet when its scale reaches the atomic scale. Interestingly, the nuclear droplet model proposed many years ago *(6)*, that is to treat the nucleus as a droplet, could successfully explain many such phenomena. These physical facts inspire us to think about the nature of the melting point of a matter. Starting from this consideration, the present paper is dedicated to present a softening theory from which one can apply it to soften the material and reduce the melting point of the mater. The basic principle lies in the tuning of the internal atomic boundary of the material which is to reduce the forces between the atoms in the matter.

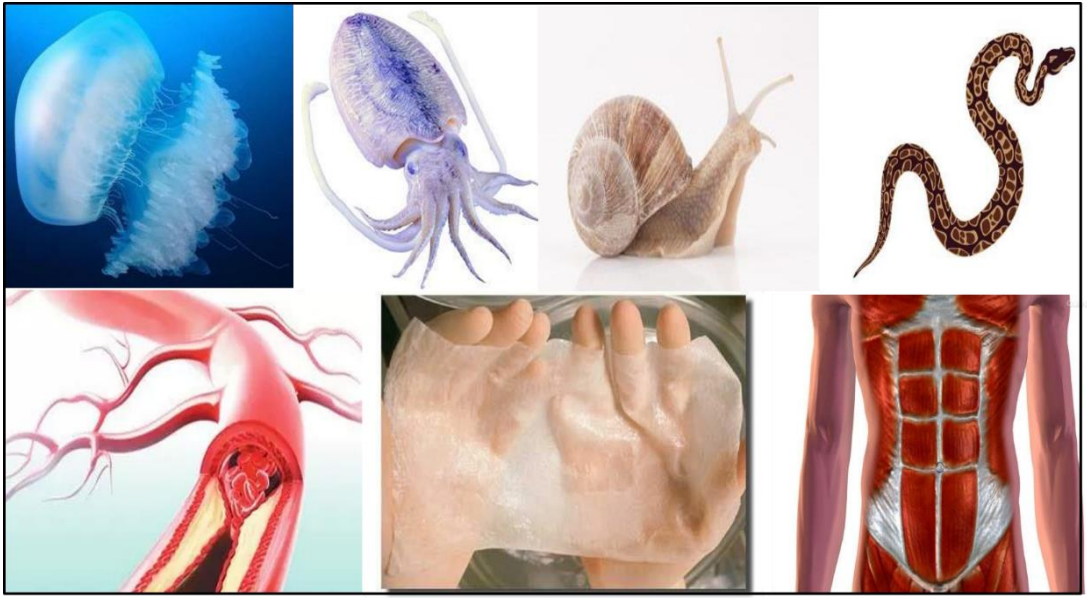

**Fig. 1.** Typical biological softness in nature and soft biological tissues in the body.

A variety of substances which have different melting points make up the colorful world. With the development of observation instruments, researchers gradually understand the micro-world including various similar atoms. To some extent, atoms which consist of nuclei and electron clouds can be considered as fluid. Of the special interest here is the case that many macroscopic substances do not remain its fluidic behavior like atoms do, that means the micro and macro consistency is destroyed. The softening theory proposed here is dedicated to revealing this phenomenon and building up a new insight for manipulating the nature in the coming time.

Softening theory which means that material bodies in the cosmic are essentially in a fluid state is illustrated as follows. Nuclei can be described as a droplet made up of protons and neutrons, while electron cloud presents the motion state of the orbital electrons (Figure 2). An actual object is composed of such fluid droplet which is a mixture of liquid and gas. Therefore, solid phase can be regarded as a temporary state for the matters indeed, while fluid phase exists permanently. Galaxies and nebulae in the cosmic perform similarly with the atomic nucleus and electrons. In this sense, all the matters can be deemed as fluid droplets with a wide variety of sizes (*7*). Softening theory introduced here provides a new interpretation on the matter.

Based on the relationship between nanoparticle size and temperature, the nuclear droplet model and alloy theory, this article expounds the connotation of softening theory. It also suggests several possible ways to tune the atomic border of matter. Finally, the article points out the limitations and challenges facing the softening theory.

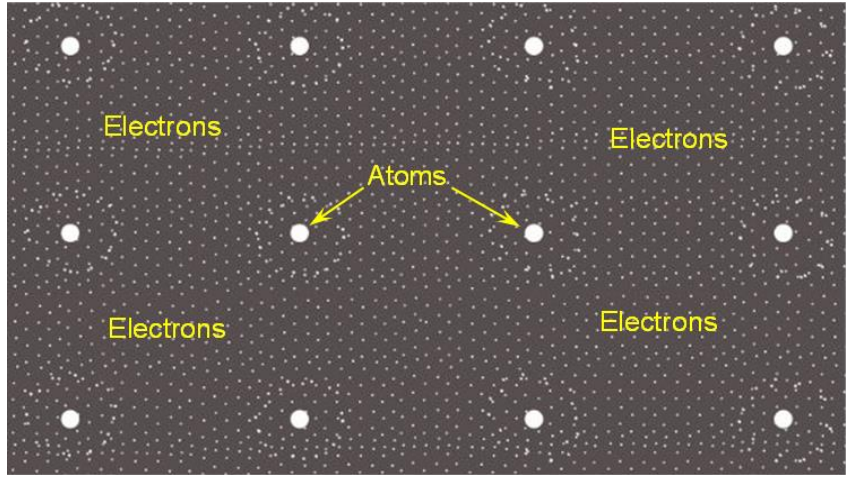

**Fig. 2.** Schematic diagram of atoms and electrons that make up matter

## 2. Basics of Softening Theory

The article proposes a basic principle to achieve the purpose of softening substances. The matter can be softened through tuning the internal atomic boundary of the matter and reducing the forces between the atoms there (as shown in Figure 3). Besides, it is equally important to prevent particles inside the matter from agglomeration.

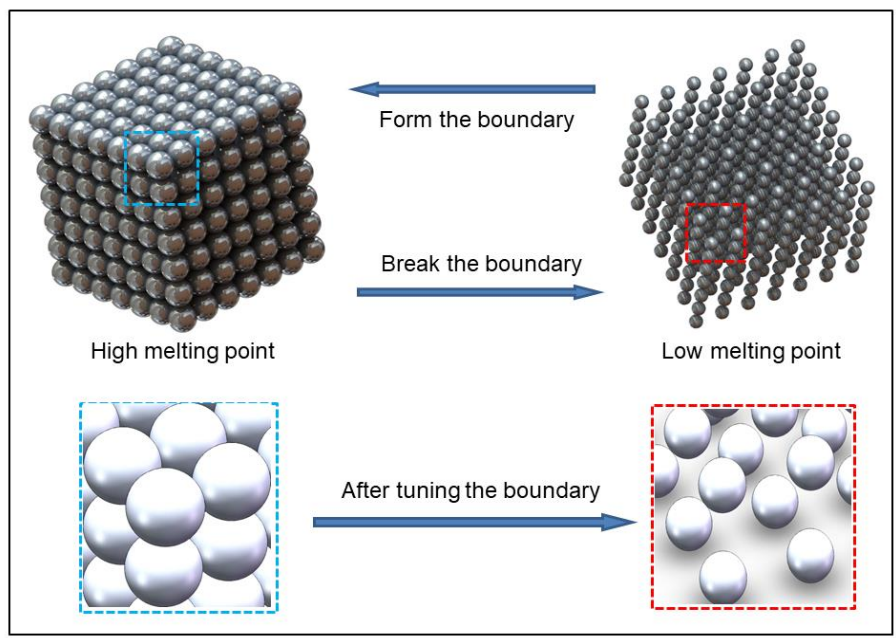

**Fig. 3.** Schematic diagram of the formation and tuning of the border.

### 2.1 Miniaturization changes the boundary to lower the melting point

Due to that small size of matter can be compared with the De Broglie wavelength of the electron, nanomaterials own a series of special physics and chemical properties such as small size effect, surface effect, quantum size effect and macroscopic quantum tunneling effect (*8*) (*9*). As the substance particles become smaller, the material's melting point is drastically reduced (*2*) (*5*) (*10*). The relationship between the melting point (T) and the radius (R) of the material is of special concern to the present theory, as the Figure 4 shows.

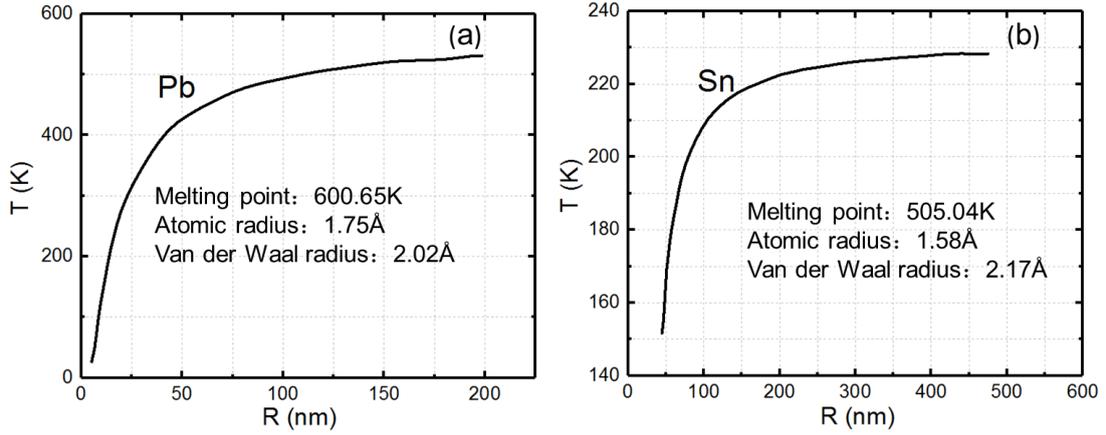

**Fig. 4.** Experimentally measured size dependence of the melting points of particles. (a) Pb particles, (b) Sn particles *(2) (5)*.

Therefore, we can draw a conclusion that there is a critical point when the size is larger than this value, the melting point remains the same, and this critical point is also generally on the nanometer scale. There are various theories for the reason of the decrease of the melting point of nanoparticles. One is described as below. The coordination number of atoms on the surface is smaller than that in the bulk, thus the interfacial energy of atoms on the surface is reduced. Therefore, the surface starts to go into disorder while the core remains in its orderly state and this is considered to be the mechanism of surface-induced melting *(9)*.

Undercooling is a very common phenomenon when liquid is solidified. The degree of undercooling is related to many factors. Generally, the degree of undercooling becomes greater with the faster cooling rate and the smaller droplet size. As the size of the droplets decreases, the droplets become more likely to remain pure with no impurities acting as crystal nuclei, which is the reason for the larger undercooling. The degree of undercooling is necessarily larger and lasts longer when the size is microscopic, which is similar to the phenomenon of the decrease of the melting point of the particles at the microscopic level. Droplets in a confined space also own a larger degree of undercooling due to the affected boundaries.

The relationship between particle size and melting point indicates that the change of the boundary of the particles greatly affects the melting point of the matter. Softening theory is committed to changing the melting point by tuning the atomic boundaries of matter.

As material particles continue to diminish, the boundaries of internal matter gradually disappear and nuclei can be seen as droplets. A well-known nuclear theory is the droplet model proposed before. Liquid droplet model, which was developed by von Weizsaecke has received extensive research and development over the past eight decades *(11) (12)*. The model presented treats the nucleus as a droplet of incompressible particle fluid. According to the model, the binding energy of a nucleus consists of volume energy, surface energy, Coulomb energy, asymmetry energy and pairing energy. And it can be expressed as:

$$B = B_v + B_S + B_C + B_a + B_p = a_v A - a_S A^{\frac{2}{3}} - a_C Z^2 A^{-\frac{1}{3}} - a_a (\frac{A}{2} - Z)^2 A^{-1} + \delta a_p A^{-\frac{1}{2}} \quad (1)$$

where $A$ and $Z$ are the number of nucleons and protons, respectively. $B_v$, $B_S$, $B_C$, $B_a$ and $B_p$ are volume energy, surface energy, Coulomb energy, asymmetry energy and pairing energy, respectively. $a_v$, $a_S$, $a_C$, $a_a$ and $a_p$ are all constants. And $\delta$ is expressed as follows,

$$\delta = \begin{cases} +1 & Z, N \text{ even} \\ 0 & A \text{ odd} \\ -1 & Z, N \text{ odd} \end{cases} \quad (2)$$

where $N$ is the number of neutrons. As $A$ becomes larger than 30, specific binding energy which is the average nuclear binding energy per nucleon remains around 8 *MeV*. It indicates that the nuclear force only exists between one nucleus and the neighboring ones. Such character is similar to that of liquid molecules. Therefore, the atom can be seen as a droplet of particle fluid wrapped by a mass of electron cloud. And the nuclear fusion and fission are also similar to the fusion and split behaviors of the water droplet.

As a conclusion, the droplet model is a kind of nuclear model established from the nature of nucleon-nucleon strong coupling in the nucleus. To a certain extent, this model can clarify the static and dynamic laws of nuclei, such as the laws of mass, surface vibrations, deformation of nuclei and nuclear fission. The atomic droplet model treats the nucleus as an ideal charged droplet and describes the nucleus kinematically according to the movement law of the droplet (*13*). Later, some new degrees of freedom have been gradually added. For example, protons and neutrons are considered as two kinds of fluids, respectively, and even different spin orientations are considered as different fluids, and compressibility and viscosity are introduced (*14*). The success of the nuclear droplet model shows the rationality of treating atoms as droplets.

## 2.2 Alloying tunes the atomic boundaries and prevents agglomeration

As a rapidly emerging class of new materials, gallium-based liquid metal as a representative of low-melting-point metal has attracted a lot of attention due to its many outstanding merits (*15-17*). Liquid metal can keep the liquid state at room temperature. As such a representative metal, the melting point of gallium is only 29.8 degrees Celsius (*18*). The singular properties of solid-state gallium and liquid gallium come from the specific arrangement of gallium atoms and electrons (*19*). The electronic arrangement is the main reason for the low melting point of gallium. Meantime, the study of gallium structure shows that there are various phases in solid gallium, where αGa is the most stable phase. There is only one nearest neighbor for each atom in αGa, a strong covalent bond between the nearest neighbors was also found (*20*) (*21*). The coexistence between covalent bonds with metal bonds at normal temperature and pressure in gallium that makes the interactions between gallium atoms smaller than the usual metal interactions, which is the reason why the melting point of gallium is low.

**Table 1**. Melting point of gallium-based alloy

| Liquid metal | Melting point /℃ |
|---|---|
| Ga | 29.8 |
| $Ga^{77}In^{23}$ | 15.7 |
| $Ga^{67}In^{20.5}Sn^{12.5}$ | 10.5 |
| $Ga^{61}In^{25}Sn^{13}Zn^{1}$ | 7.6 |
| $Ga^{67.98}In^{20.01}Sn^{10.5}Ag^{1.51}$ | <4 |

At the same time, researchers noticed that the melting point of the alloy is lower than any melting point of the component materials in many cases (Table 1). For example, the gallium indium binary alloy which is shown in Figure 4 has the lowest melting point of 15 degrees Celsius. The reason behind this is that the doping between different metals tunes the atomic boundary of the original metals. The alloys are divided into two types, solid solutions and compounds. The alloy compounds refer to the formation of new materials in the alloy (*22*). The solid solutions refer to the mutual replacement of atoms at a certain crystalline structure of the metal without changing the structure and symmetry of the entire crystal (*23*). For solid solutions, doping between different alloys can change the boundaries between the interior of the metal, thereby reducing the melting point of the solid solution (Figure 5).

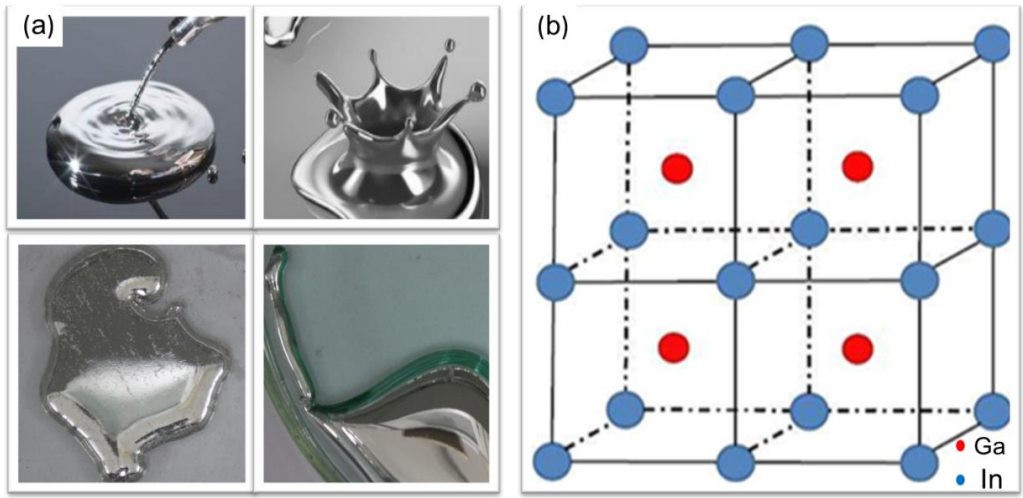

**Fig. 5.** (a) Ga-In alloy with excellent softness. (b) Structural model of eutectic Ga-In alloy.

Overall, softening theory for matter commits to figure out possible ways to regulate the melting point of substances to achieve the desired soft materials. From the above discussion, it can be found that the rationality and success of the theory did get supported via the existing experimental evidences although still limited to some extent. For example, alloy theory indicates that the melting point of the matter can be completely changed by tuning the internal boundaries of matter through mixing different matching metals together. Clearly, many soft materials can be possibly made out along the category of softening theory.

3. **Potential Softening Methods**
   As raised above, tuning the atomic boundary is vital for softening matter. However, regulating borders requires appropriate energy injection. In this aspect, external energy which can act in atomic level is critically required in order to achieve the desired softening goal. Based on the proposed principles, several potential ways to achieve material softening can be proposed as below (as shown in Figure 6). In principle, these methods are devoted to tuning the atomic boundaries of matter and preventing the re-aggregation of matter.

3.1. **Mechanical force**
   Mechanical dispersion refers to the use of mechanical force to prevent particle from agglomeration. It is necessary that the mechanical force (referring to fluid shear and compressive stress) should be larger than the adhesion between particles. Generally, the source of mechanical force can be selected as the intense turbulent flow of air caused by the jet and impact of a high-speed rotating impeller disk or high-speed air stream.

3.2. **Electric field**
   Electrostatic repulsion can play a role due to the same surface charged for homogeneous particles. Therefore, the electrostatic force can be utilized for particle dispersion. The key issue is the way to make the particle group fully charged. Contact charging, induction charging and other approaches can make the particles charged, but the most effective way is to charge the corona, the continuous supply of particles through the corona discharge to form an iron curtain.

### 3.3. Ultrasonic wave

The particles can fully disperse through dealing with the material directly placed in the ultrasound field and controlling the appropriate ultrasonic frequency and duration of action. Ultrasonic dispersion is the use of ultrasonic cavitation generated by local high temperature, high pressure, strong shock waves and micro-jet, etc. To significantly weaken the interaction between the particles can effectively prevent the particle from agglomeration and being fully dispersed.

### 3.4. Magnetic field

A field refers to a spatial area in which objects of a certain nature exert a force on similar objects that do not touch them. Through the role of the applied field, it can effectively regulate the force of the particles in order to achieve the effect of softening substances. Naturally, the electric and magnetic field, thermal field and other fields can be administrated.

### 3.5. Optics

Based on the energy equation of light, the energy of light is related to optical frequency. As the optical frequency increases, photon energy becomes larger. Hence, the atomic boundary of matter can be broken by using the optics, leading to the lower melting point to soften the matter. Besides, one can regulate the atomic boundary of matter and then realize different degrees of softening through selecting optical beams with different frequencies.

### 3.6. Heat

It is well known that the matter will melt when the heating temperature reaches its melting point. The reason behind this is that heat can provide energy which is used for the break of boundary. Therefore, the atomic boundary can be tuned by heating.

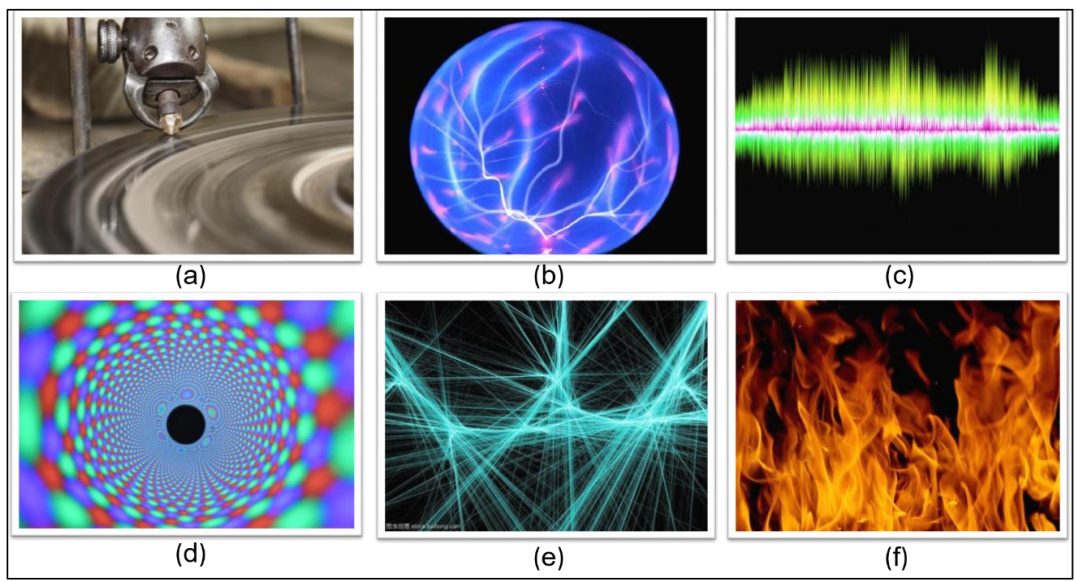

**Fig. 6.** Potential softening methods. (a) Mechanical force. (b) Electric field. (c) Ultrasonic wave. (d) Magnetic field. (e) Optical beams. (f) Heat.

### 4. Discussion

In current article, a theory of softening materials has been proposed and several potential ways to achieve softening matter are also discussed. Overall, such softening theory may offer an unconventional interpretation on the melting point of different matters. It implies theoretical significance for regulation of softness and melting point of the substance. After all, the theory is still in its infancy stage, there are some issues worth further justifications.

**Quantification of softening degree.** Softening theory can guide the way to reduce the melting point of the matter. The degree of reduction should be able to use equations to describe. In the future, more theoretical research on softening theory should be carried out.

**Experimental evidences.** Softening theory for matter introduced here needs more experiments to justify its rationality and adaptability.

**Effective softening methods.** Softening theory devoted to tuning the melting point of the matter requires more effective softening techniques due to tremendous energy required during the material treatment process.

**Philosophical thinking.** The structure of the lattice may subject to change when the forces between atoms which make up a particular substance are removed completely or partially. Therefore, it is worth considering whether the matter has become another different one.

## 5. Conclusion

In summary, the softening theory for matter which is committed to regulating the softness and melting point of the substance is proposed in this article. The matter can possibly be softened through tuning the internal atomic boundary of the matter which thus reduces the forces between the atoms in the matter. Softening theory which means that material bodies in the cosmic are essentially in a fluid state is preliminarily illustrated. An atomic nucleus can be described as a droplet made up of protons and neutrons, while electron cloud presents the motion state of the orbital electrons. An actual object is composed of such fluid droplet which is a mixture of liquid and gas. Therefore, solid phase can be regarded as a temporary state for the matters, while fluid phase exists permanently. Galaxies and nebulae in the cosmic perform similarly to the atomic nucleus and electrons. In this sense, all the matters can be deemed as fluid droplets with a wide variety of sizes. Overall, the softening theory as proposed here contributes to the understanding of the material behavior and may help build up a new insight for understanding the matters in nature.

**Acknowledgment**

This work was partially supported by the NSFC Grant 91748206, Funding of Higher Education Agency and Frontier Funding of Chinese Academy of Sciences.